\documentclass[a4paper,11pt]{article}
\usepackage{jinstpub} 
\usepackage{lineno}
\usepackage{siunitx}
\usepackage{graphicx}
\usepackage{subfigure}
\usepackage[version=4]{mhchem}

\newcommand{\hso}{\ce{H2SO4}}

\title{\boldmath Titanium for rare-event searches: Hydrofluoric acid-free etching}

\author[a,1]{P. Knights,\note{Corresponding author.}}
\author[a,b]{K. Nikolopoulos,}
\author[a]{G. Rogers,}
\author[a,2]{D. Spathara,\note{Corresponding author.}}
\author[a]{P. Walters}
\affiliation[a]{School of Physics and Astronomy, University of Birmingham, Birmingham, B15 2TT, UK}
\affiliation[b]{Institute for Experimental Physics, University of Hamburg, Hamburg, 22761, Germany}

\emailAdd{p.r.knights@bham.ac.uk}
\emailAdd{d.spathara@bham.ac.uk}

\abstract{Rare-event search experiments require construction materials with high radiopurity to minimise background contributions. Thanks to its high mechanical strength, low density, machinability, and commercial availability in relatively radio-pure forms, titanium is a suitable material for structural elements in rare event searches. To remove surface deposits on materials used, a chemical etching stage is usually performed. However, the chemical resistance of titanium means that, conventionally, such etching is done with hydrofluoric acid. Hydrofluoric acid presents serious health risks to users, and such hazards are compounded in the case of construction in deep underground laboratories. An alternative chemical etching using sulphuric acid is presented. This is demonstrated to etch titanium, removing $3.7\,\si{\micro\meter}$ of material from the surface over the course of 20~hours. Scanning electron microscopy with back-scattered electron spectroscopy was used to study the surface and contamination of the titanium, demonstrating the removal of surface contaminants after etching. The proposed method is a potential alternative to those currently employed.}


\begin{document}
\maketitle
\flushbottom

\section{Introduction}
Rare-event search experiments are key to understanding  dark matter and neutrinos. Given that these experiments are searching for some of the rarest processes, controlling experimental backgrounds is of paramount importance. As such, they require unprecedented levels of radiopurity in the materials from which they are constructed. This has driven extensive research into the development of radiopure materials, e.g.~\cite{HOPPE2007486,Hoppe:2014nva, Vitale:2021xrm,  Knights:2025ogz, Spathara:2025hrp, Spathara:2025bfw}. Where high-strength is required, such as for pressure vessels, Ti is a candidate material. For example, Ti was utilised by the LUX-ZEPLIN experiment~\cite{LZ:2015kxe, LZ:2024zvo}, and is a potential material for future experiments, e.g. the nEXO~\cite{nEXO:2017nam} and XLZD~\cite{XLZD:2024nsu}. 

While its high tensile strength, machinability, low density, and availability in high purities are clear strengths, Ti has some limitations, such as it cannot be found commercially at purities comparable to those of copper. Moreover, during the production of all materials, some of the tooling material or dust from the environment can become deposited in the surface layers of the material. As a result, progeny of radioactive elements in deposited dust or of gaseous $^{222}$Rn present in the environment, can implant into the surface of the material. 

To mitigate such backgrounds, often construction materials are subsequently coated with higher purity layers~\cite{NEWS-G:2020fhm}. This requires chemical preparation of the surface, performed through chemical etching, preferably after all mechanical assembly is completed. The high chemical resistance of Ti makes this challenging and conventional methods employ hydrofluoric acid (HF)-based solutions or proprietary mixtures. While proprietary mixtures are not preferable due to the uncertainty on the ingredients, which themselves could cause radioactive contamination, the use of HF is especially challenging as it poses serious health risks. HF is classed as acutely toxic, and carries European Regulation (EC) No 1272/2008 hazard statements H300, H310, and H330 (Fatal if swallowed, in contact with skin or if inhaled). As such, extensive controls and specialist training is required for its use. These risks are further accentuated when used in deep underground laboratories. As a result, a simpler -- both in terms of procedure and chemical mixture -- alternative etching method would be transformative to our ability to use Ti in high-radiopurity applications. 

Another use of Ti is in medical settings, for example, for dental implants, internal fixation plates, and prosthetics. In Ref.~\cite{BAN20061115} 48\% sulphuric acid (\hso), heated to a range of temperatures, was studied for the surface modification of commercially pure Ti. \hso~at mixtures of a few percent is already used in deep-underground laboratories for processes such as high-purity copper electroplating. Thus, an investigation of the \hso~concentrations for Ti etching is particularly relevant.

In the presented work, Ti etching with 40\% \hso~at $40\,^{\circ}$C has been studied. 

\section{Etching Method}
A sample of grade 1 Ti (commercially pure, unalloyed) were prepared into a $3\,\si{\centi\meter}$ in diameter, $1\,\si{\centi\meter}$ thick disk. This was cleaned in ethanol in an ultra-sonic bath at room temperature temperature for $80\;\si{minutes}$. After cleaning, the mass of the sample was $28.056\,\si{\gram}$.  

A solution of 40\% \hso~in type 2 water ($1\,\si{\mega\ohm\,\centi\meter}$) was prepared in a cleaned glass beaker. The beaker was then placed
in a water bath set at $40\,^{\circ}$C and left for $15\,\si{minutes}$ to equilibrate. The Ti sample was then placed into the beaker and completely submerged in the solution. No sonication or agitation of the solution was used. The Ti was left to dwell in the solution for 20~hours. During this time, the mean temperature of the bath was $32\,^{\circ}$C. 

After etching, the sample was removed from the solution and thoroughly rinsed with type 2 water. The sample was transferred to an oven to be dried at $50\,^{\circ}$C for 5 hours. Subsequently, the mass of the sample was recorded as $28.031\,\si{\gram}$. Assuming uniform removal on the whole surface, this corresponds to an etched thickness of $3.7\,\si{\micro\meter}$.

\section{Sample Characterization}
The surface of the sample was studied before and after etching using Scanning Electron Microscopy (SEM) in Secondary Electron Imaging (SEI) and Backscattered Electron Imaging (BEI) modes, for topology and composition investigations, respectively. Composition profiles were measured using Energy-Dispersive X-ray (EDX) spectroscopy. These were performed with the JEOL 7000F field emission gun–scanning electron microscope at an accelerating voltage of $20\;\si{\kilo\volt}$ and a working distance of $10\;\si{\milli\meter}$. EDX data are collected with a silicon-drift X-ray detector. The surface was also studied macroscopically, using the KEYENCE VHX-7000 Digital Microscope, both for its topology and roughness.

\begin{figure*}[h]
    \centering
    \subfigure[\label{fig:unEtchedSEI}]{\includegraphics[width=0.48\linewidth]{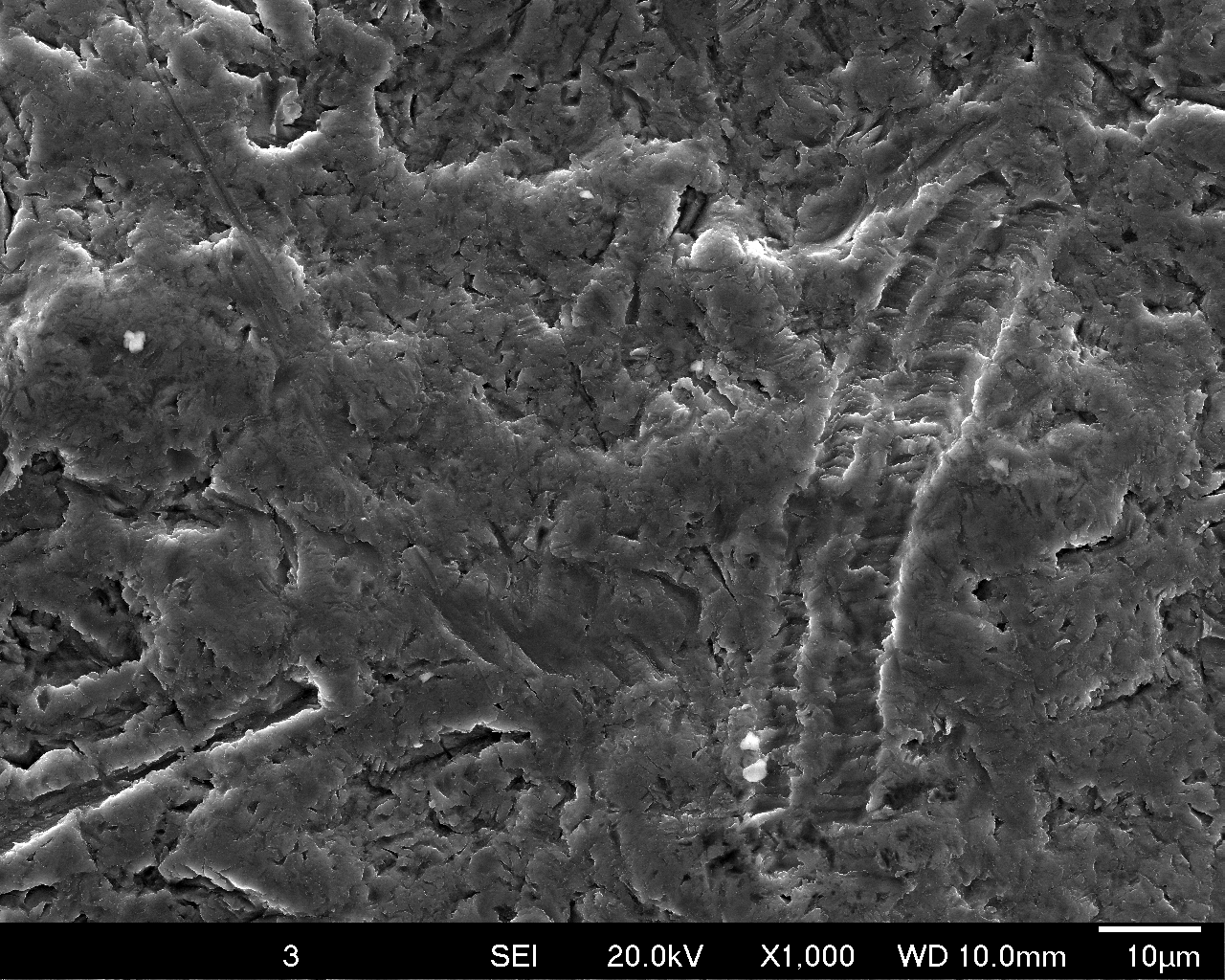}}
    \subfigure[\label{fig:unEtchedBSE}]{\includegraphics[width=0.48\linewidth]{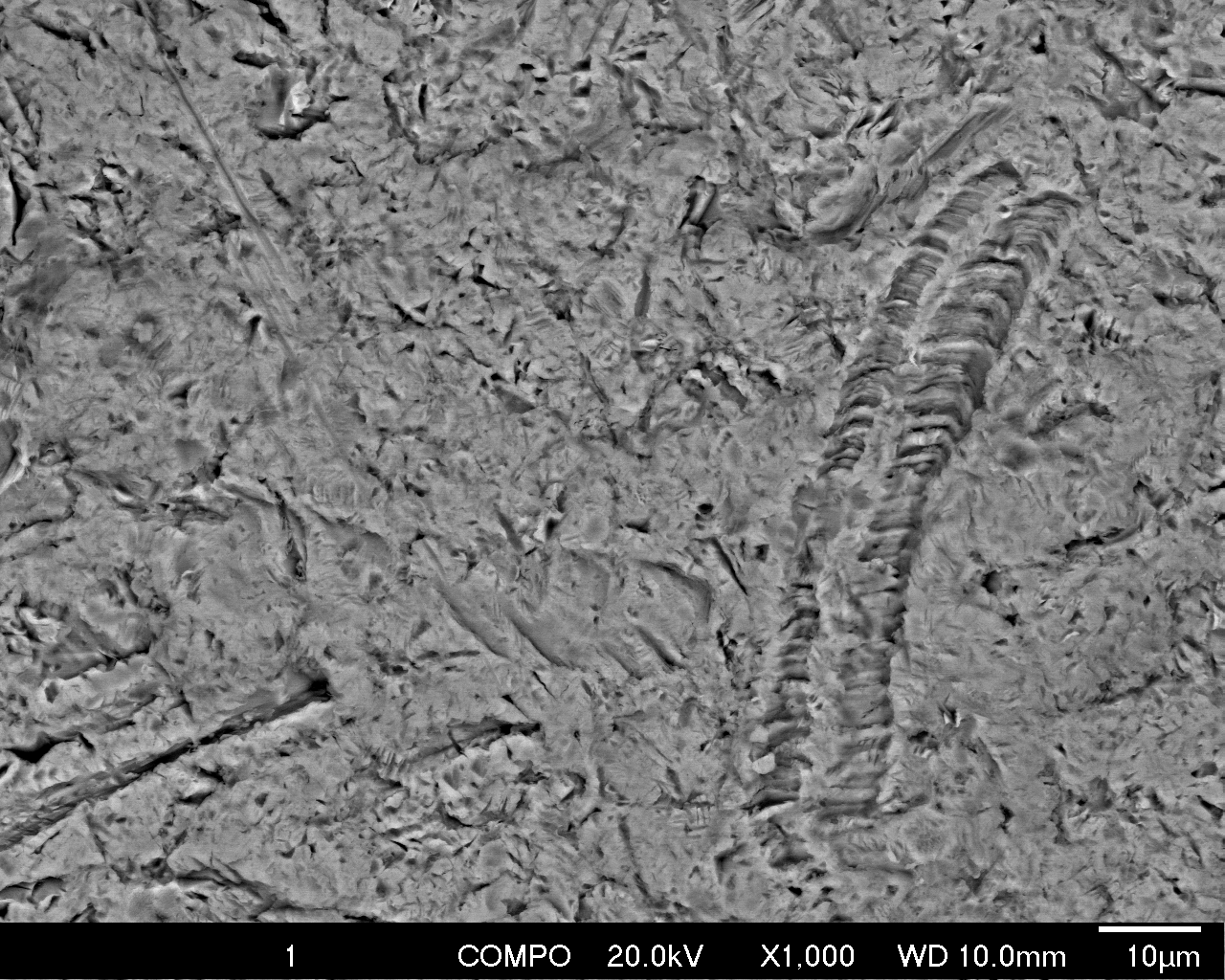}}
    \subfigure[\label{fig:etchedSEI}]{\includegraphics[width=0.48\linewidth]{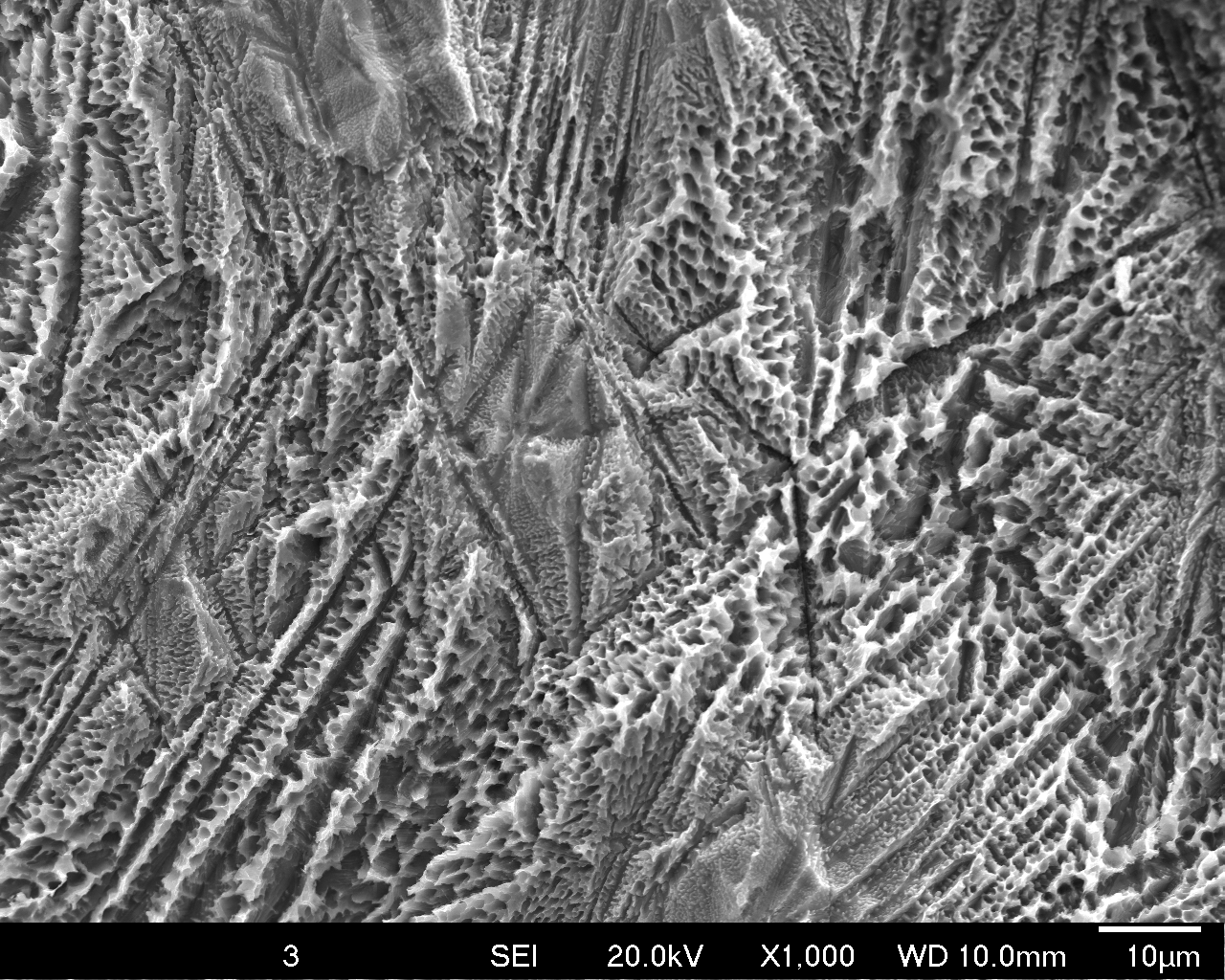}}
    \subfigure[\label{fig:etchedBSE}]{\includegraphics[width=0.48\linewidth]{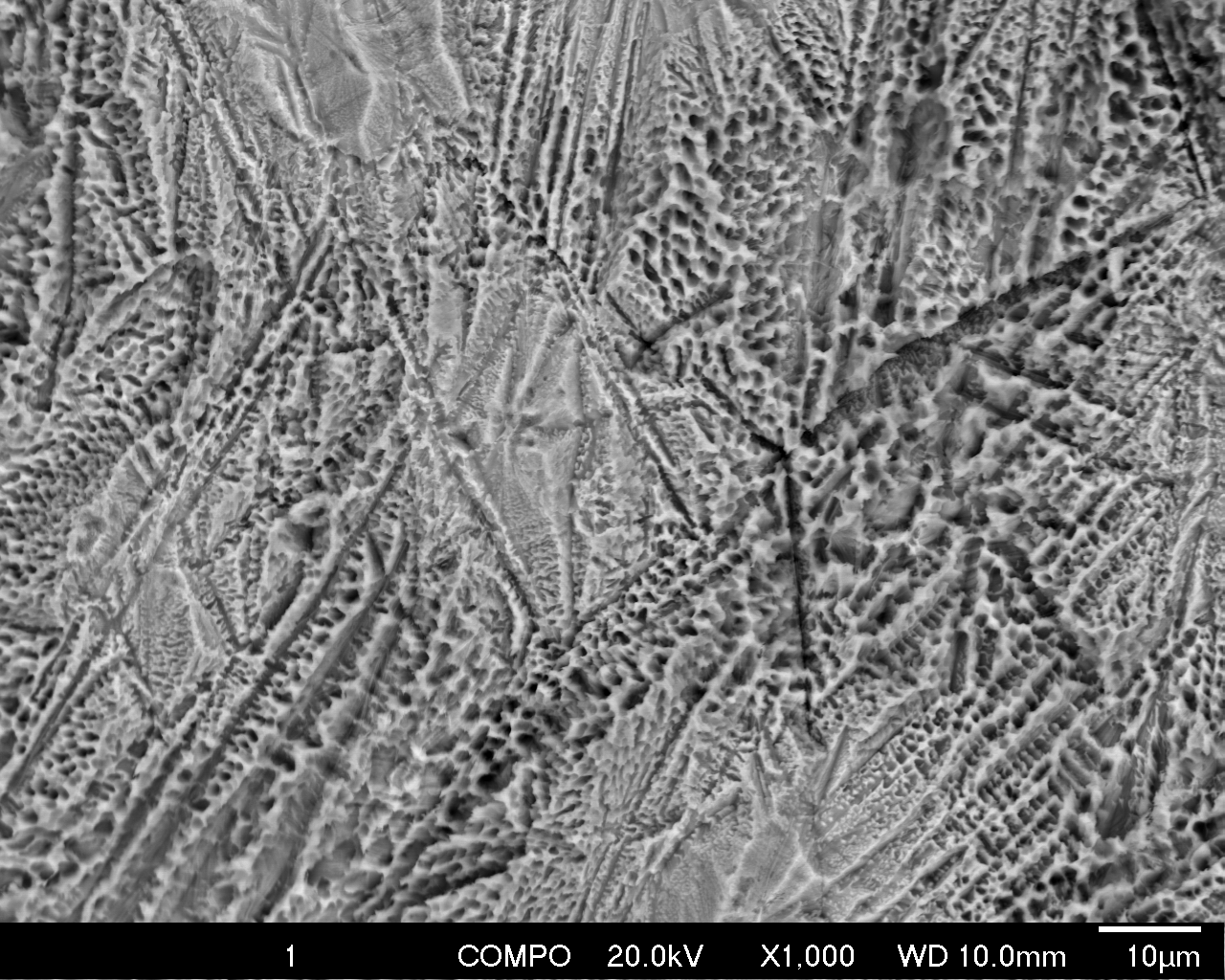}}
    \caption{Imaging of Ti surface using SEM JEOL 7000F with x1000 magnification: pre-etching sample \subref{fig:unEtchedSEI} electron image (SEI), \subref{fig:unEtchedBSE} backscattered image (BSE); and etched sample \subref{fig:etchedSEI} electron image (SEI), \subref{fig:etchedBSE} backscattered image (BSE).
    \label{fig:SEIandBSE}}
    \vspace{-0.3cm}
\end{figure*}

Figure~\ref{fig:unEtchedSEI} and \subref{fig:unEtchedBSE} show the results of the SEM imaging of the Ti surface before etching in SEI and BSE modes, respectively. 
The corresponding imaging studies of etched Ti are shown in Figures~\ref{fig:etchedSEI} and \subref{fig:etchedBSE}, for the SEI and BSE modes, respectively. 

The image of the surface pre-etching shows some irregular topology in the SEI mode. However, the topology after etching reveals the dendritic structure beneath the initial surface of the sample, while the dendritic arm spacing is also visible. The visibility of dendritic structures, which form during the manufacturing process, highlights the removal of the irregular surface topology during etching.

\begin{figure*}[h!]
    \centering
    \subfigure[\label{fig:2a}]{\includegraphics[width=0.48\linewidth]{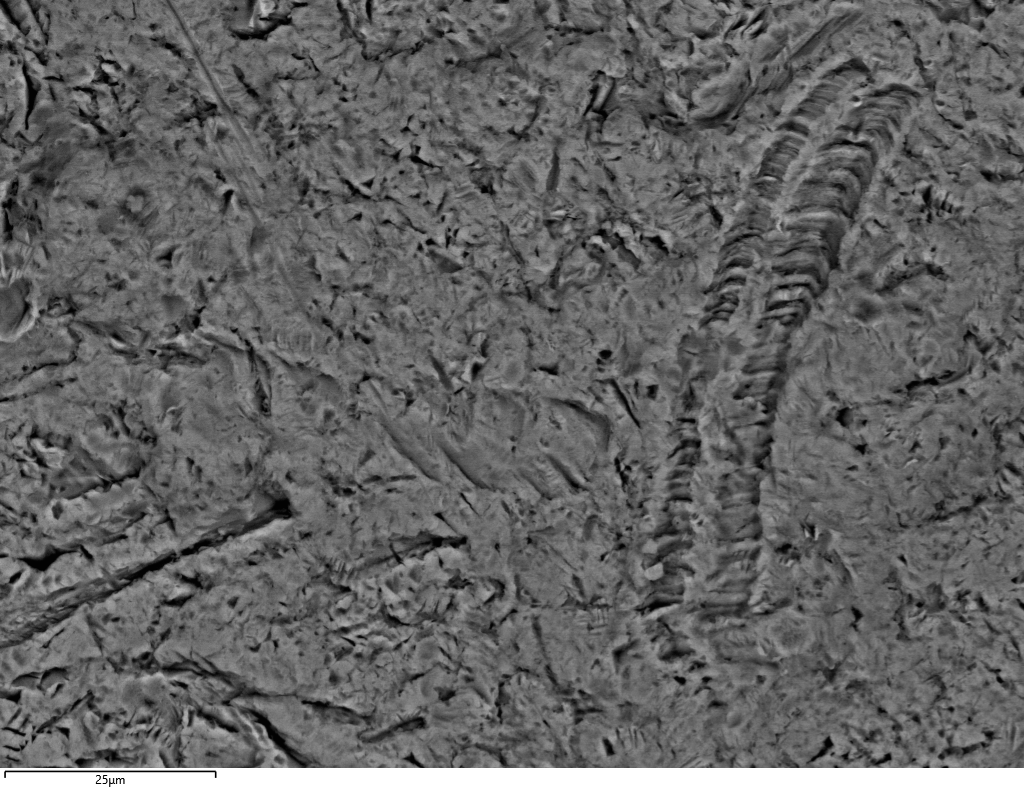}}
    \subfigure[\label{fig:2b}]{\includegraphics[width=0.48\linewidth]{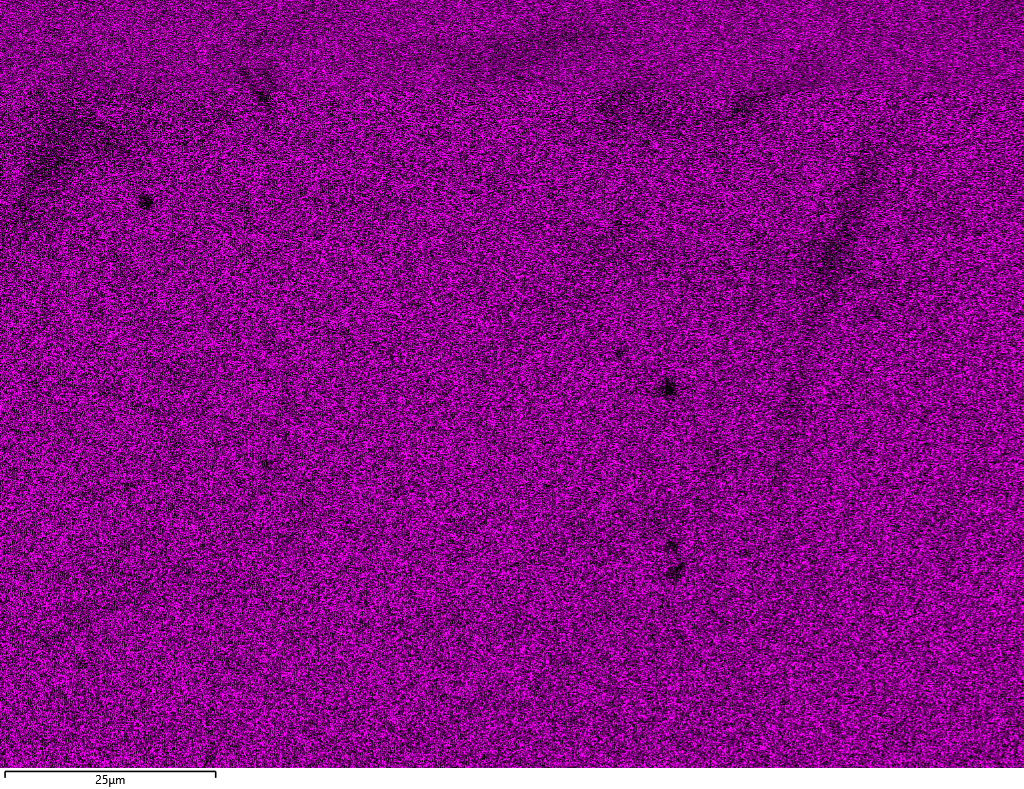}}
    \subfigure[\label{fig:2c}]{\includegraphics[width=0.48\linewidth]{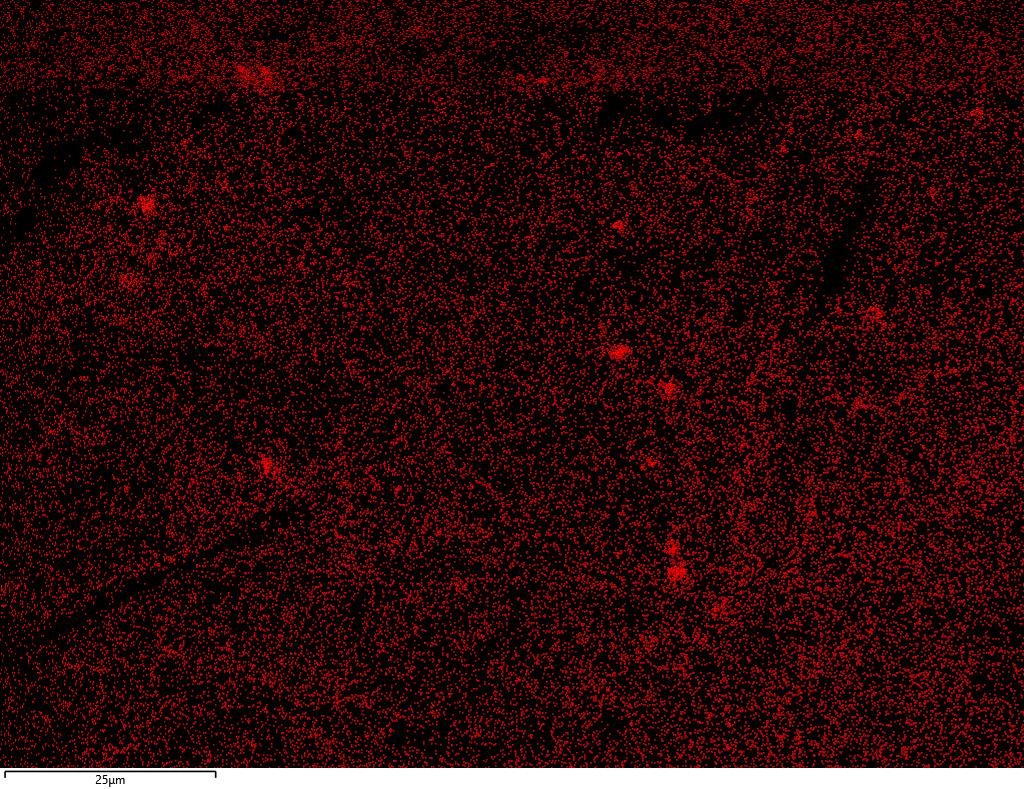}}
    \subfigure[\label{fig:2d}]{\includegraphics[width=0.48\linewidth]{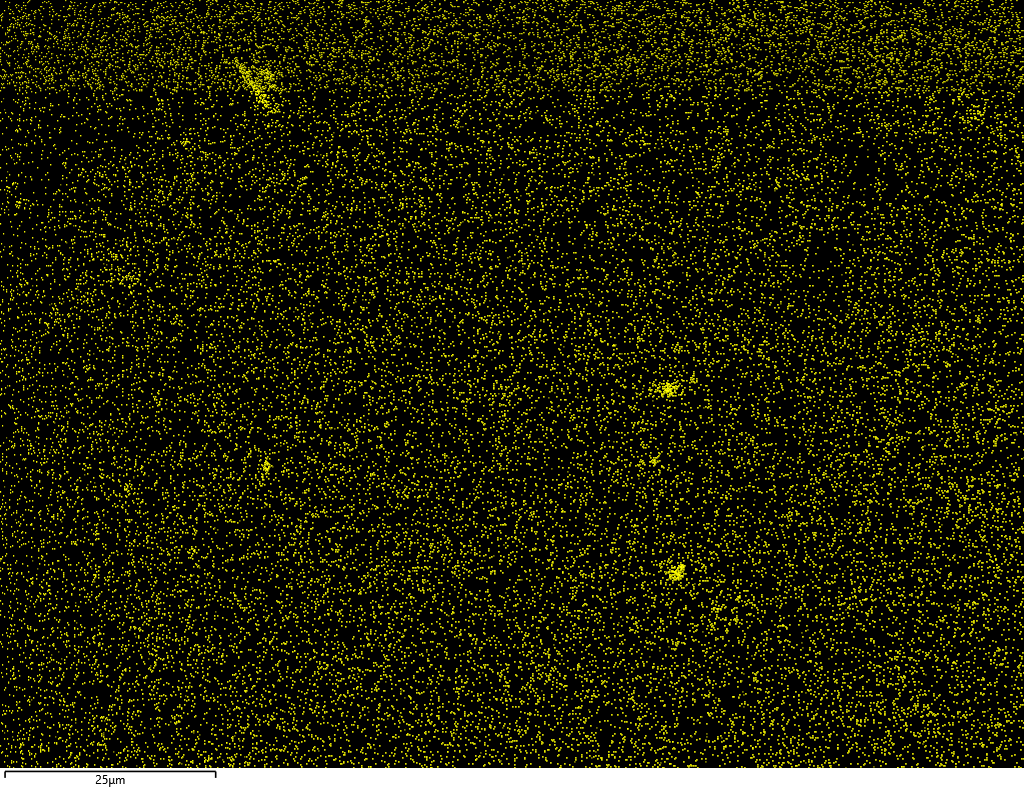}}
    \subfigure[\label{fig:2e}]{\includegraphics[width=0.48\linewidth]{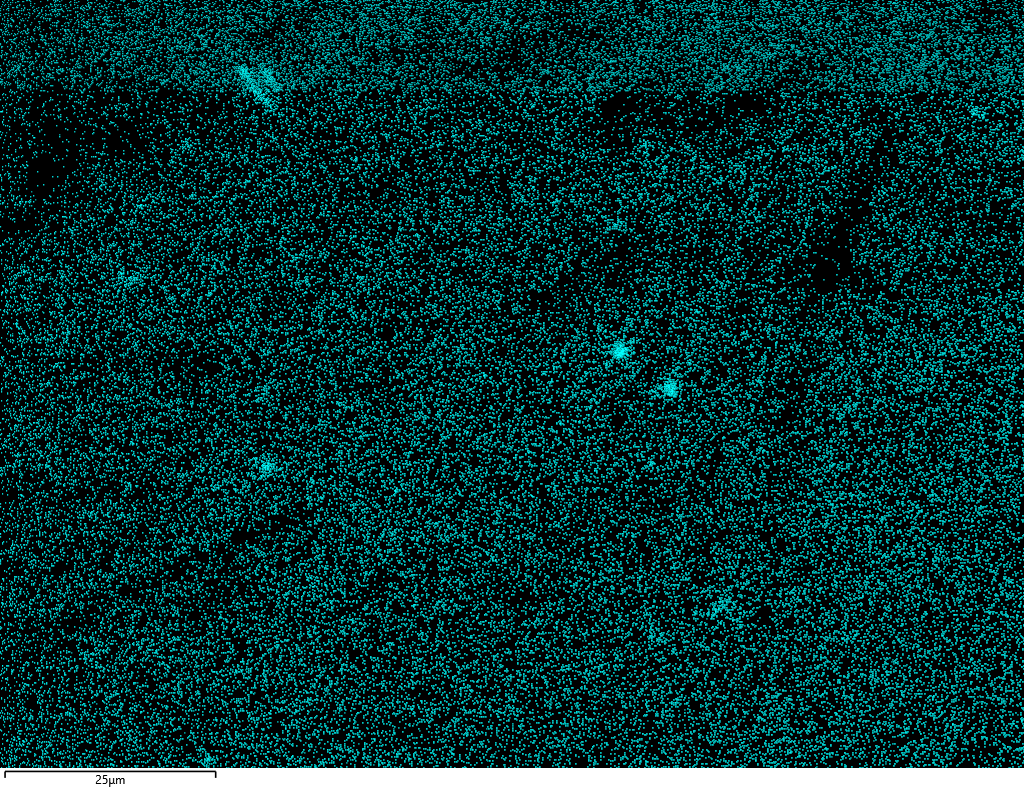}}
    \caption{\subref{fig:2a} BSE image of Ti surface prior to etching and the EDX maps of the surface showing the atomic contributions from \subref{fig:2b} Ti, \subref{fig:2c} O, \subref{fig:2d} Fe, and \subref{fig:2e} Si. Brighter colours indicate a higher atomic concentration. 
    \label{fig:2}}
    \vspace{-0.3cm}
\end{figure*}

\begin{figure*}[h]
    \centering
    \subfigure[\label{fig:3a}]{\includegraphics[width=0.48\linewidth]{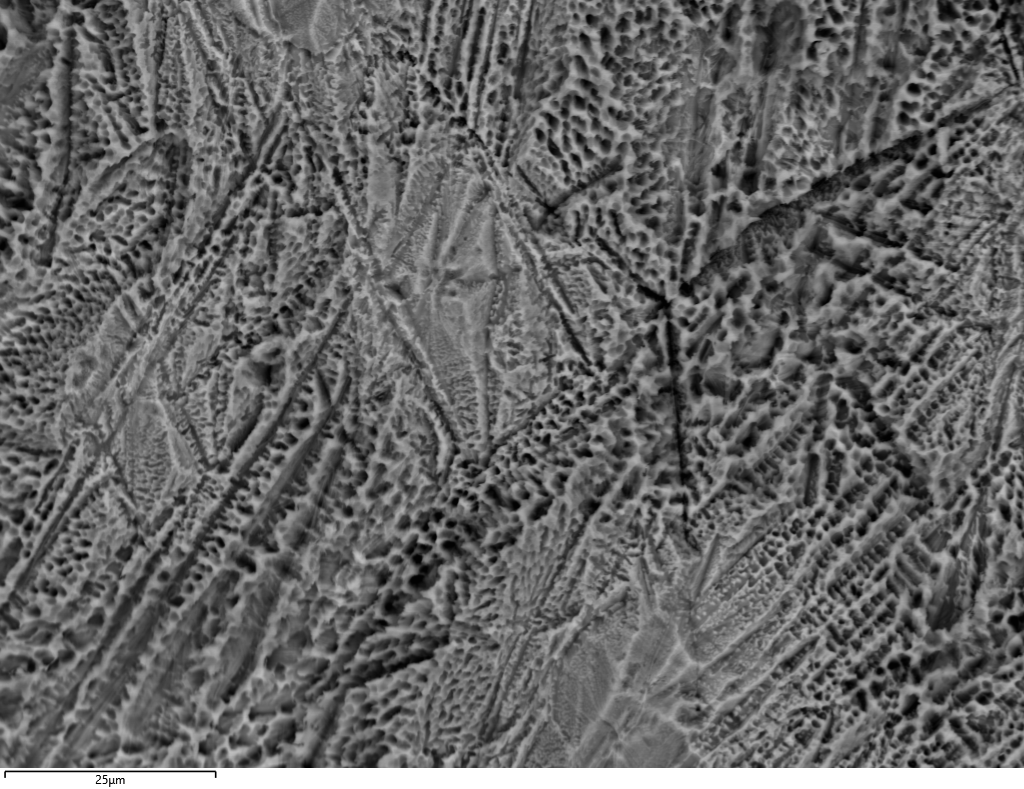}}
    \subfigure[\label{fig:3b}]{\includegraphics[width=0.48\linewidth]{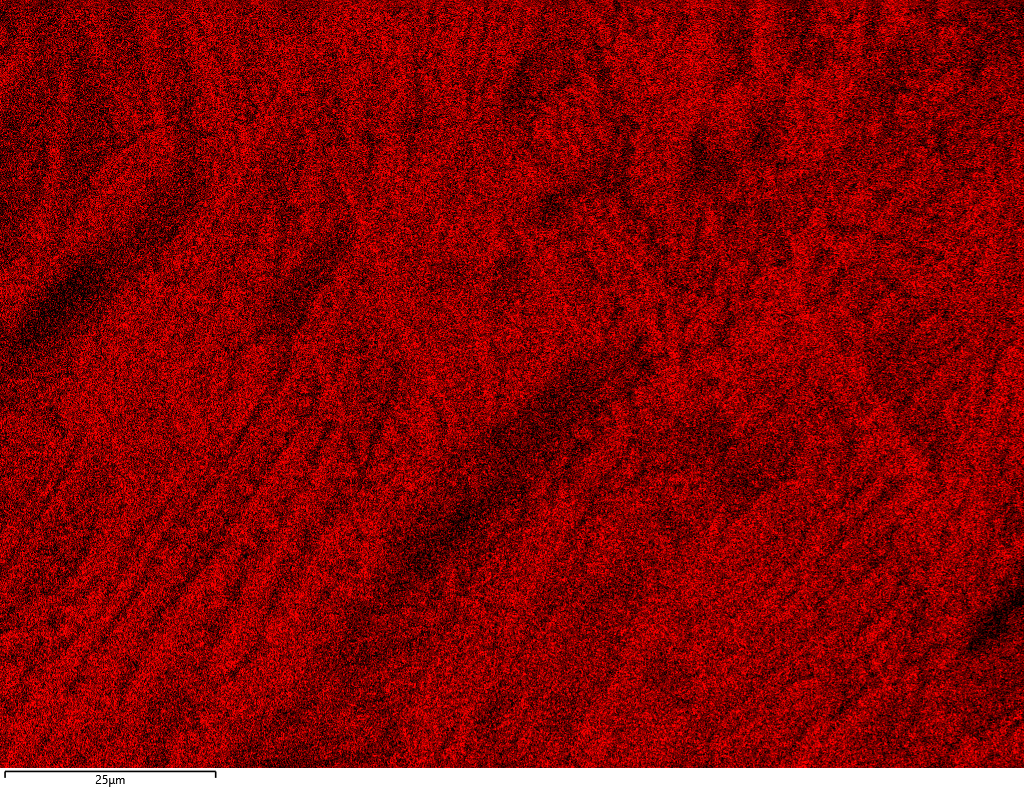}}
    \caption{\subref{fig:3a} BSE image and \subref{fig:3b} EDX maps of etched Ti surface using SEM JEOL 7000F.
    \label{fig:3}}
    \vspace{-0.3cm}
\end{figure*}

The image in BSE mode before etching does not show any significant contrast. However, it is possible to locate spots with higher concentration in heavier elements by comparing with the EDX maps, shown in Figure~\ref{fig:2}. Thus, the reason why the contrast is not high in the BSE mode image is that these spots contain both lighter and heavier than Ti elements in significant concentrations. Conversely, the image in BSE mode after etching does not show any significant contrast either, because as shown in Figure~\ref{fig:3} the surface comprises practically only Ti.

The BSE mode was also used to study the atomic composition of the surface, both on EDX maps, shown in Figure~\ref{fig:2}, and using EDX point measurements. This highlighted locally significant oxygen, iron, and silicon contamination. 
It is noted that the Ti surface is not prepared using metallography (i.e. grinding/polishing) and, 
as a result, point measurements for contamination cannot be obtained for samples which are not adequately levelled due to roughness.
Nevertheless, the X-ray detector used to collect EDX data is capable to deliver high-fidelity images of local contaminations, providing confidence in the observed local high contamination in O, Fe, and Si.

Figure~\ref{fig:3} shows a region of the Ti sample after etching. EDX point measurements indicate a Ti concentration greater than 98\%. 
Since the surface is not polished, the areas where etching revealed the dendritic structure appear with higher contrast due to their relative depth rather than indicating a somewhat lower Ti concentration in the map. Given that all apparent depletions in Ti observed in the EDX measurement are matched by the dendritic structure, this indicates that contamination of the Ti surface is eliminated after etching and a pure Ti surface is obtained.

\begin{figure*}[h]
    \centering
    \subfigure[\label{fig:1a}]{\includegraphics[width=0.48\linewidth]{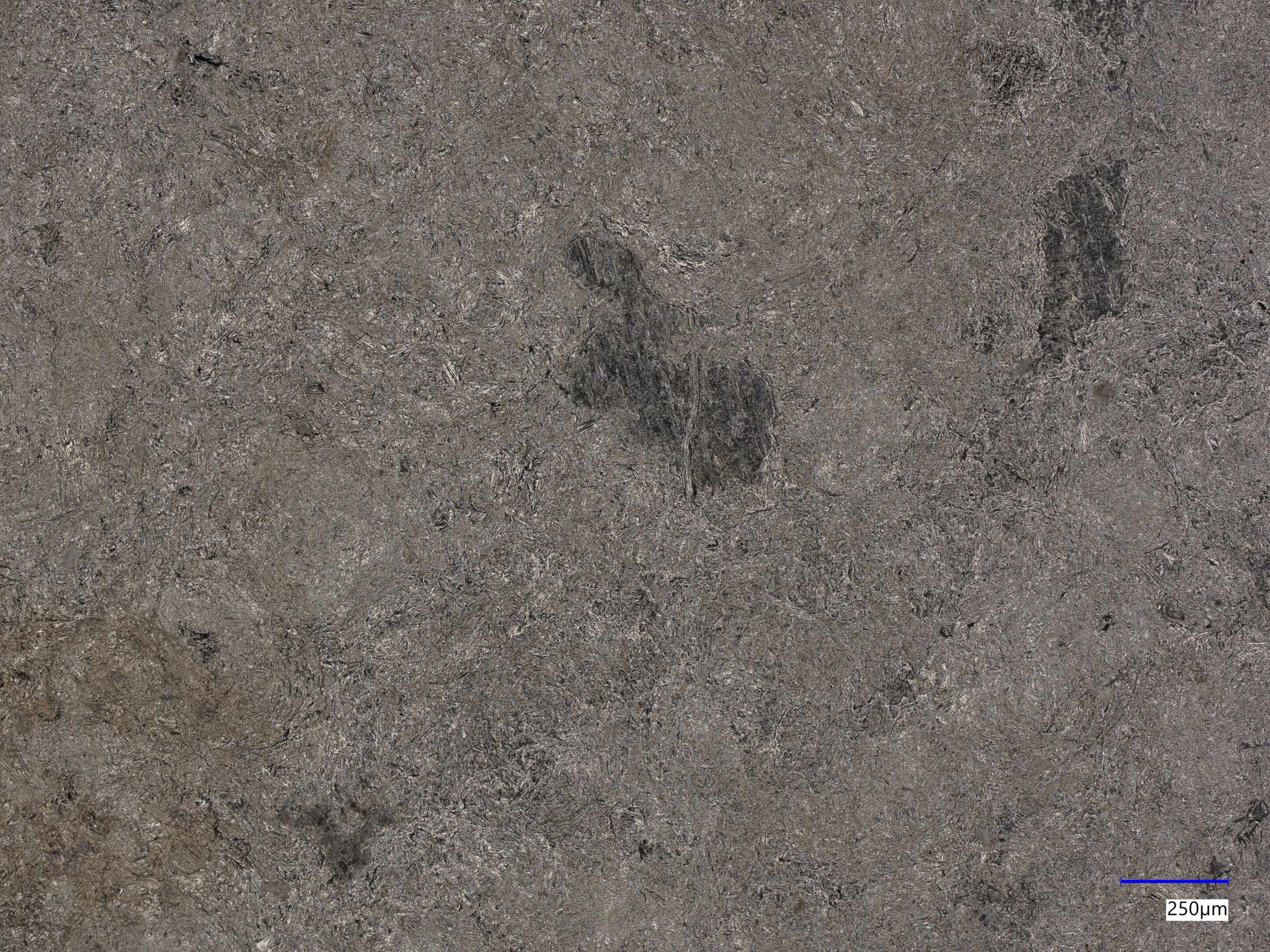}}
    \subfigure[\label{fig:1b}]{\includegraphics[width=0.48\linewidth]{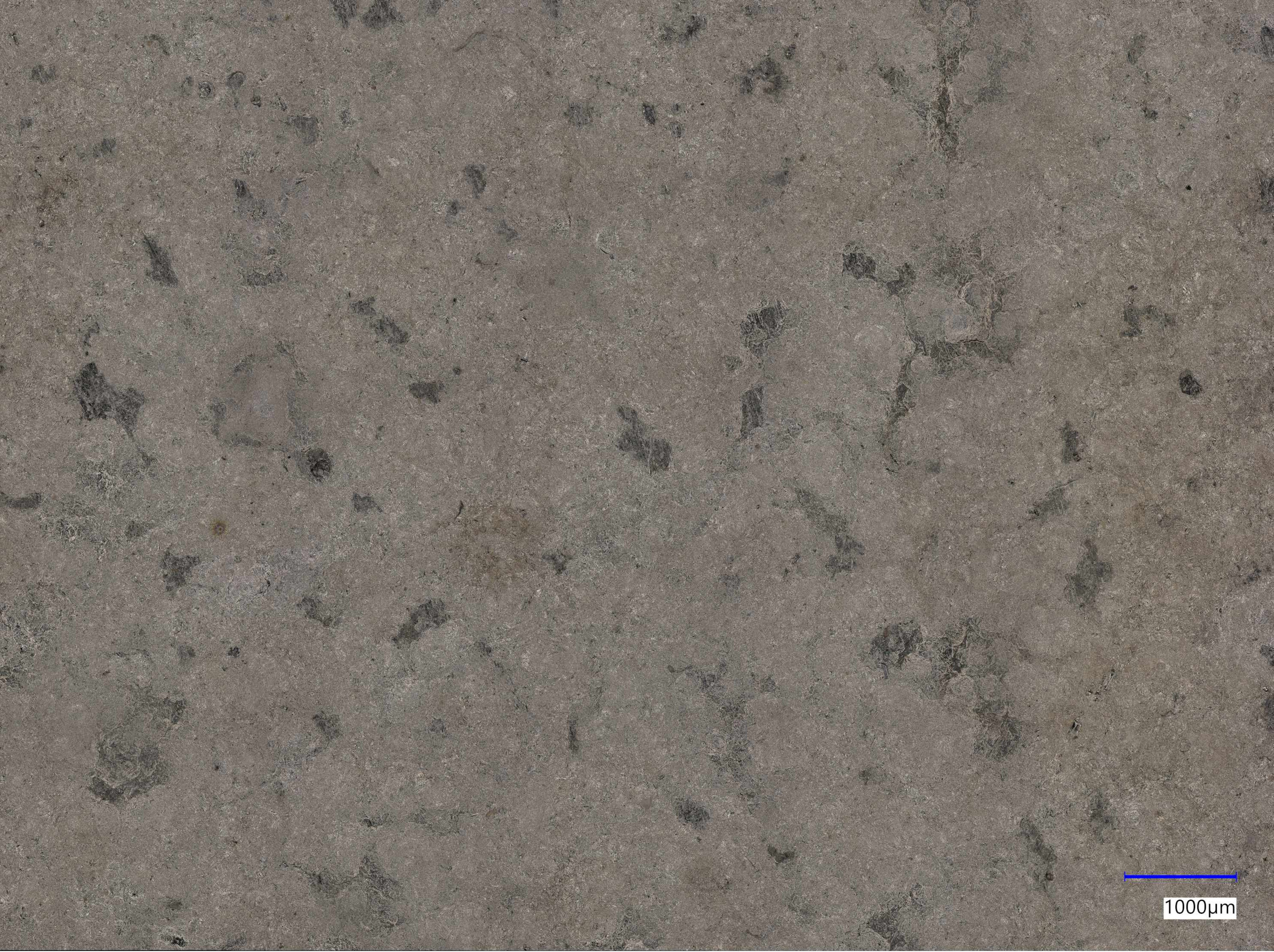}}
    \subfigure[\label{fig:1c}]{\includegraphics[width=0.48\linewidth]{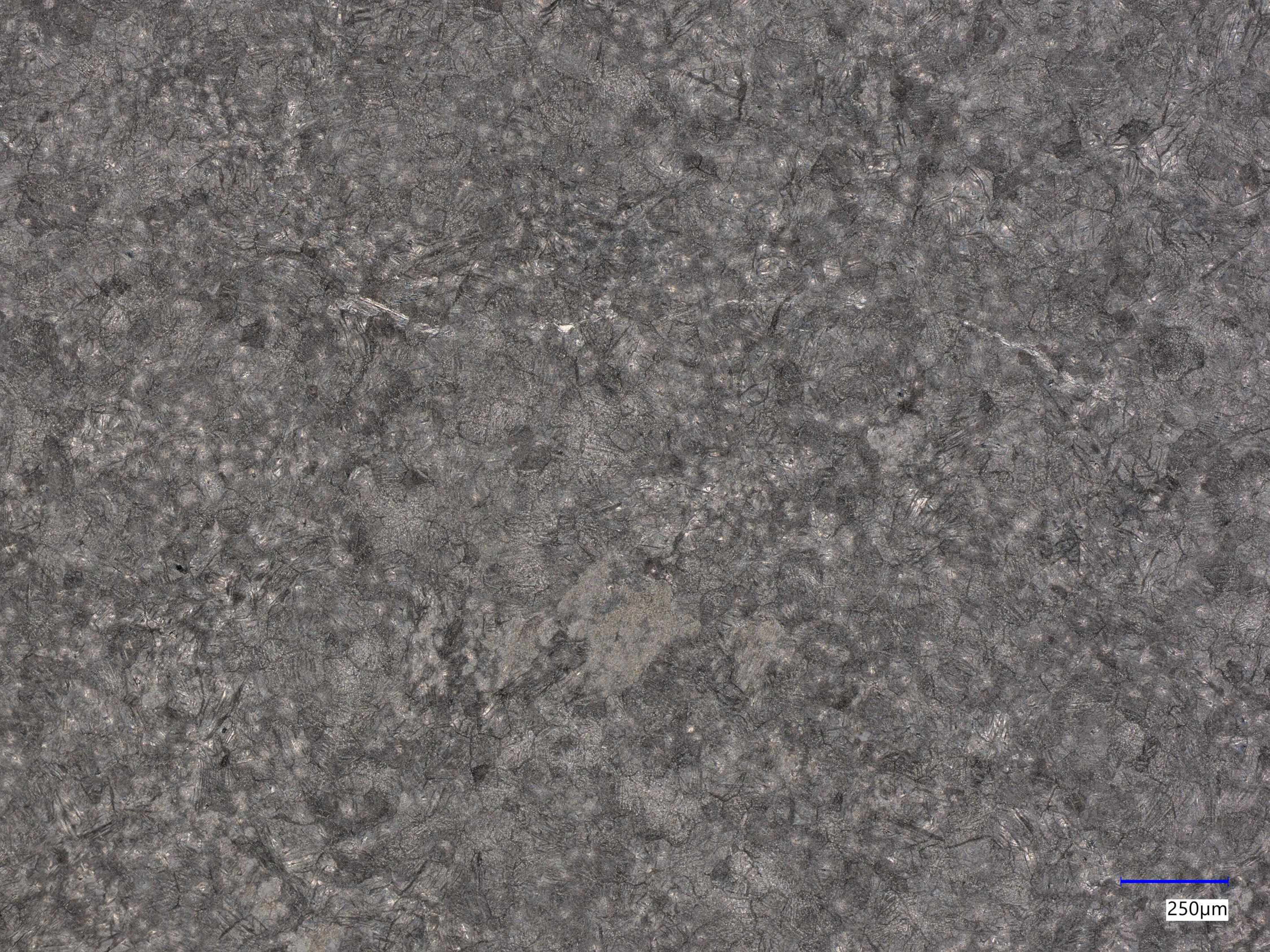}}
    \subfigure[\label{fig:1d}]{\includegraphics[width=0.48\linewidth]{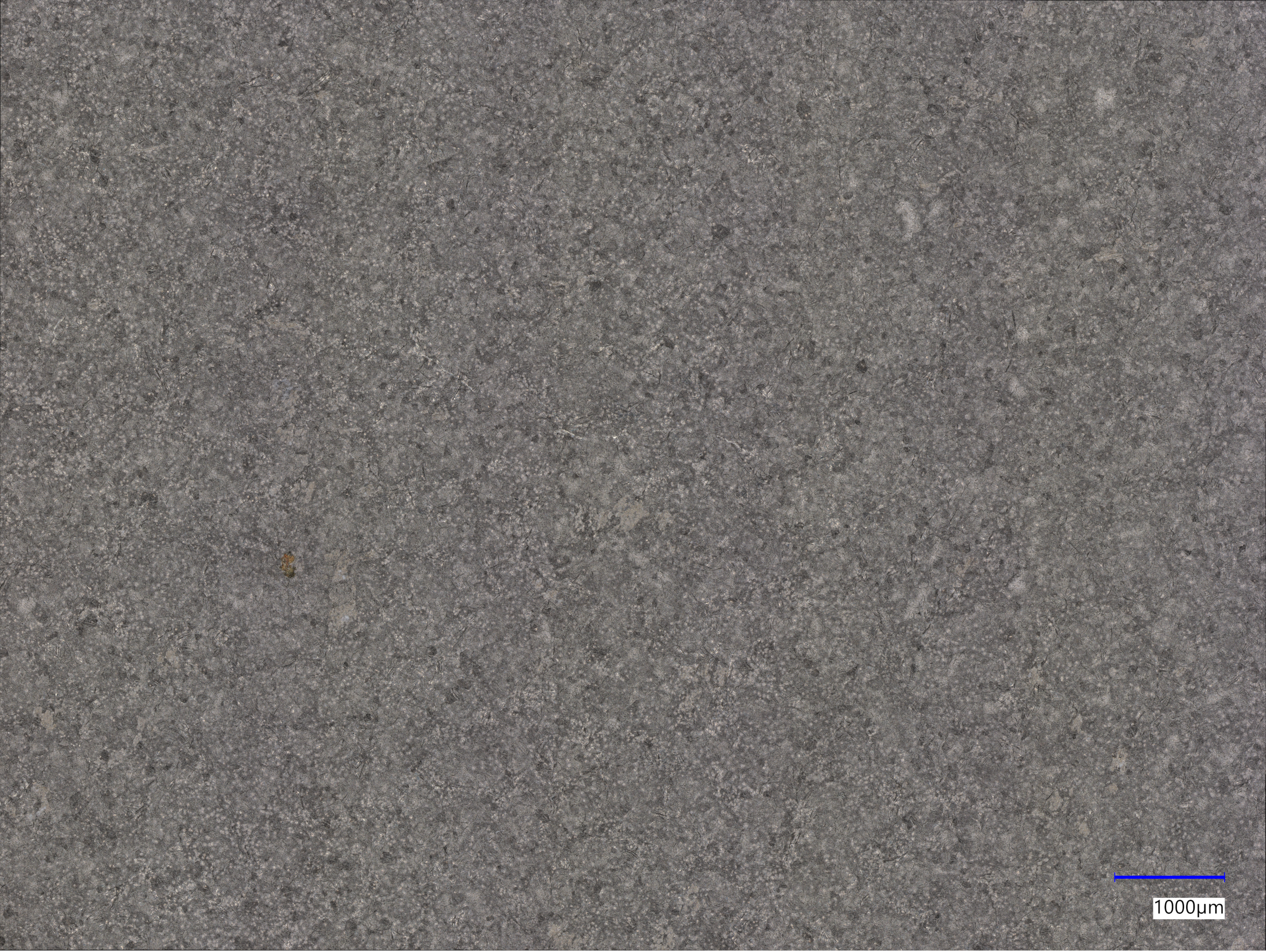}}
    \caption{Imaging using the KEYENCE VHX-7000 digital microscope of Ti surface \subref{fig:1a} x100 pre-etching,  \subref{fig:1b} 5x5 frames of x100 pre-etching, \subref{fig:1c} x100 etched, 
    \subref{fig:1d} 5x5 frames of x100 etched.
    \label{fig:1}}
    \vspace{-0.3cm}
\end{figure*}

Figures~\ref{fig:1a} and \subref{fig:1c} show the pre-etching and etched surfaces, respectively, using the KEYENCE VHX-7000 digital microscope at x100 magnification. Extended areas, where 5x5 overlapping frames each of x100 magnification have been stitched using the 3D mode and are depicted in Figures~\ref{fig:1b} and \subref{fig:1d}. This allows for improved focus of each frame independently of its local roughness.  
The images of the pre-etching Ti surface show dispersed valleys compared to the images of the etched surface, where irregularities are smaller and uniformly distributed. This indicates the increased surface uniformity achieved by the etching procedure. 

\section{Summary}
A hydrofluoric acid-free etching method for Ti is presented using sulphuric acid. Thanks to the simplicity of the etchant, comprising \hso~and water, the components are significantly safer and easier to work with than HF, and can also be easily screened for radioactivity, an important consideration for rare-event search experiments. The ability to etch Ti at relatively low concentrations of \hso~further enhances its applicability. SEM measurements of Ti before and after etching demonstrated the removal of persistent surface contaminants, not removed by simple ethanol sonication.

\acknowledgments

Support from the UKRI-STFC (No. ST/W000652/1, No. ST/Y509036/1, No. ST/Z000777/1) and the UKRI Horizon Europe Guarantee scheme (PureAlloys – EP/X022773/1) is acknowledged. The support of the Deutsche Forschungsgemeinschaft (DFG, German Research Foundation) under Germany’s Excellence Strategy – EXC 2121 “Quantum Universe”-390833306 is acknowledged. SEM, EDX, studies were performed at the Electron Microscopy Group Facility, University of Birmingham.

\bibliographystyle{JHEP}
\bibliography{biblio.bib}

\end{document}